\documentclass[5p]{elsarticle}

\usepackage{lineno}
\modulolinenumbers[5]

\journal{Physica E}

\usepackage{amsmath}
\usepackage{amssymb}
\usepackage{amsopn}
\usepackage{graphicx}
\usepackage{dcolumn}
\usepackage{bm}
\usepackage{ulem}
\usepackage{epsfig}
\usepackage{color}
\usepackage{textcomp}
\usepackage{comment}
\usepackage[lofdepth,lotdepth,caption=false]{subfig}
\usepackage[percent]{overpic}
\usepackage[breaklinks=true,colorlinks,citecolor=blue,linkcolor=blue,urlcolor=blue]{hyperref}

\DeclareMathOperator{\e}{e}

\begin{document}
\title{Finite-frequency noise in a topological superconducting wire}

\author{Stefano Valentini}
\address{NEST, Scuola Normale Superiore, and Istituto Nanoscienze-CNR, I-56126 Pisa, Italy}

\author{Michele Governale}
\address{MacDiarmid Institute for Advanced Materials and Nanotechnology, School of Chemical and Physical Sciences, Victoria University of Wellington, P.O. Box 600, Wellington 6140, New Zealand}

\author{Rosario Fazio}
\address{NEST, Scuola Normale Superiore, and Istituto Nanoscienze-CNR, I-56126 Pisa, Italy}

\author{Fabio Taddei \corref{mycorrespondingauthor}}
\address{NEST, Istituto Nanoscienze-CNR and Scuola Normale Superiore, I-56126 Pisa, Italy}
\cortext[mycorrespondingauthor]{Corresponding author}
\ead{fabio.taddei@sns.it}

\begin{abstract}
In this paper we study the finite-frequency  current cross-correlations for a topological superconducting nanowire attached to two terminals  at one of its ends. Using an analytic 1D model we show that the presence of a Majorana bound state yields vanishing cross-correlations for frequencies larger than twice the applied transport voltage, in contrast to what is found for a zero-energy ordinary Andreev bound state. Zero cross-correlations at high frequency have been confirmed using a more realistic tight-binding model for finite-width topological superconducting nanowires. Finite-temperature effects have also been investigated.

\noindent \,

\noindent 
Contribution for the special issue of Physica E in memory of Markus B\"uttiker.
\end{abstract}

\maketitle

\section{Introduction}
%

One of the prototypical systems which host Majorana Bound States (MBS) is the Kitaev chain\cite{Kitaev2001}, a discrete model  for a one-dimensional $p$-wave superconductor.
Such a model can be realised in a semiconducting nanowire with strong spin-orbit coupling by  placing it in close proximity to a $s$-wave superconductor, thus inducing superconductivity in the wire,  and applying a strong magnetic field which leads to a large Zeeman splitting\cite{Lutchyn2010,Oreg2010}.
With possible solid-state realisations available, several experimental studies have gathered evidence compatible with the  existence of MBSs\cite{mourik2012,das2012,deng2012,finck2013,churchill2013}.
 As a result, the quest for an unambiguous signature of the presence of a MBS has become a priority and is stimulating a significant research effort\cite{alicea2012,leijnse2012,beenakker2013,Stanescu2013, Dassarma2015}. 
Up to now only a few papers  have focused on the consequences of MBSs on the behaviour of current correlations \cite{Golub2011,Wu2012,Lu2012,Wang2013,Liu2013,Zocher2013,Lu2014,Soller2014,Li2014,Gnezdilov2015,Ulrich2015}.
Very recently Haim et al. (Ref.~\cite{Haim2015}) have considered the spin-resolved current cross-correlations, finding that they are negative for a MBS in the case of correlations between opposite spins.

\begin{figure}[t]
\centering
\includegraphics[width=0.9\columnwidth]{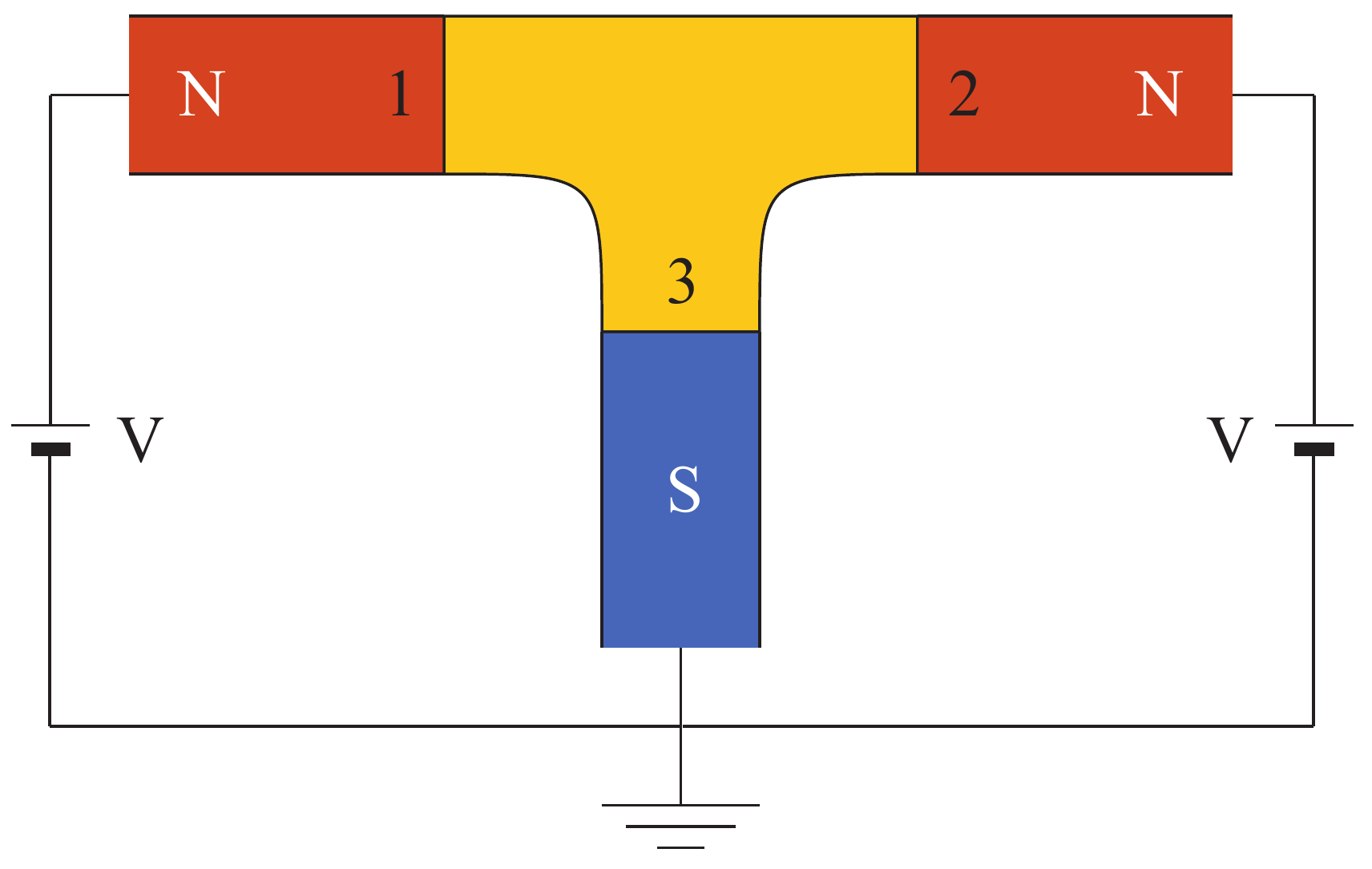}
\caption{Schematic setup of the system. A superconducting (S) nanowire, which can be driven in and out of  the topological phase by an applied magnetic field, is contacted by means of a beam splitter to two normal (N) leads, labelled 1 and 2.}
\label{setup}
\end{figure}

In this paper, we consider a semi-infinite topological superconducting wire attached to two normal terminals at one end, as shown in Fig.~\ref{setup}.
A bias voltage $V$ is applied to the two normal contacts (labelled 1 and 2), while the superconducting wire is grounded.
We calculate the cross-correlations at finite frequency between the currents $I_1$ and $I_2$ flowing in the two normal leads.
Our main finding is that the cross-correlations at frequencies larger than twice the voltage $V$ vanish at zero temperature, when the superconducting wire is in the topological phase.
On the contrary, in the presence of an ordinary zero-energy Andreev Bound State (ABS), which similarly to a MBS gives rise to a zero-bias peak in the differential conductance, the cross-correlations at high frequency are in general non-zero and depend on the details of the system.
The origin of this phenomenon can be attributed to the peculiar structure of the energy-dependent scattering matrix for reflections off a MBS.
In addition, at zero-frequency the cross-correlations are always negative in the presence of a MBS, analogously to the case of spin-resolved spin-up/spin-down cross-correlations\cite{Haim2015}, but may be positive in the case of an ABS.
These results have been obtained with a simple analytical model whereby two normal leads are coupled either to a Majorana state or a zero-energy level.
Zero cross-correlations at high frequency are then confirmed using a more realistic tight-binding model based on the semiconducting-nanowire realisation of a one-dimensional $p$-wave superconducting wire.

\section{Formalism and methods}

In this section we briefly review the scattering approach for the finite-frequency current-current correlations in hybrid superconducting systems \cite{Anantram1996}.
The (non-symmetrized) current-current correlator between lead $i$ and $i'$ is defined as
\begin{equation}
\label{noisespin}
S_{ii'}(t)=\langle   I_{i}(t)  I_{i'}(0) \rangle - \langle   I_{i} \rangle \langle  I_{i'} \rangle ,
\end{equation}
where $ I_{i}(t)$ is the current%
\footnote{In this paper we consider only quasiparticles current and neglect the role of displacement currents; the latter might induce corrections to the noise at high frequency in the case of strongly energy-dependent density of state \cite{Buttiker1992,Blanter2000,Pedersen1998}. This must be taken into account in the analysis of actual experimental data. Such corrections, however, depend on the details of the system and their discussion is beyond the scope of the present paper.} 
operator at time $t$ relative to terminal $i$ and $\langle \cdots \rangle$ stands for the quantum-statistical average.
Taking the Fourier transform of $S_{ii'}(t)$ one obtains the finite-frequency correlator as
\begin{equation}
S_{ii'}(\omega)=\int S_{ii'}(t) \; e^{i\omega t} \; dt .
\end{equation}
Within the Landauer-B\"uttiker scattering approach\cite{Blanter2000,Dattabook}, the finite-frequency current-current correlator in a hybrid superconducting system, calculated in the  normal leads, is given by\cite{Anantram1996}
\begin{align}
\label{Sspin}
S_{ii'}(\omega)&=\frac{e^2}{2h} \sum_{\substack{\alpha,\alpha' \\ \beta,\beta'}}
\sum_{\substack{\sigma,\sigma' \\ \tau,\tau'}}
\sum_{j,j'}
\text{sign} (\beta) \text{sign} (\beta')\\
& \times \int_{-\infty}^{+\infty} dE \; A^{\alpha j \sigma}_{\alpha' j' \sigma'} (\beta,E,E+\hbar\omega,\tau,i) \nonumber \\
& \times A^{\alpha' j' \sigma'}_{\alpha j \sigma} (\beta',E+\hbar\omega,E,\tau',i') \nonumber \\
& \times f_{\alpha i}(E)\left[ 1-f_{\alpha' i'}(E+\hbar\omega) \right],  \nonumber
\end{align}
where the indices $\alpha,\alpha',\beta,\beta'\in\{\pm1\}$ indicate electrons (+1) and holes (-1) in Nambu space, $\sigma,\sigma',\tau,\tau'$  refer to the  spin-projection quantum number and $i,i',j,j'$ label  the leads.
The Fermi distribution function in the normal lead $i$ for a $\alpha$-like particle at temperature $T$ and voltage $V_i$  is given by
\begin{equation}
f_{\alpha i}(E)=\left[ 1+\exp \left(\frac{E-\alpha eV_i}{k_BT}\right)\right]^{-1} .
\end{equation}
Moreover, in Eq.~(\ref{Sspin}) we have defined
\begin{align}
\label{A}
 A^{\alpha j \sigma}_{\alpha' j' \sigma'} (\beta,E,E',\tau,i)&=
\delta_{\alpha\beta} \delta_{\sigma\tau} \delta_{ji} \delta_{\alpha'\beta} \delta_{\sigma'\tau} \delta_{j'i}   \nonumber \\
& -[s^{i \tau j \sigma}_{\beta \alpha}(E)]^\star s^{i \tau j' \sigma'}_{\beta \alpha' }(E') ,
\end{align}
where $s^{i \tau j \sigma}_{\beta \alpha} (E)$ is the scattering amplitude at energy $E$ for a $\alpha$-like particle with spin $\sigma$ injected from lead $j$ to be reflected as a $\beta$-like particle with spin $\tau$ in lead $i$. 
In the rest of this paper, we shall focus on the symmetrized noise, defined as $S^S_{ii'}(\omega)=S_{ii'}(\omega)+S_{ii'}(-\omega)$, since this is the quantity that is measured by a classical detector \cite{Nazarov2009}. 
We shall furthermore assume that the two normal terminals are kept at the same voltage ($V_1=V_2=V$).

\section{Analytic 1D model}
\label{a1d}
The system depicted in Fig.~\ref{setup} can be modelled in a simple way by composing\cite{Dattabook} the scattering matrix $s^{\text{M}}$ of a normal lead coupled to a Majorana state with the scattering matrix $s^{\text{bs}}$ of a 3-leg beam splitter, which describes the connection to terminals 1 and 2.
The matrix $s^{\text{M}}$ can be calculated from the Hamiltonian describing a normal lead coupled to a Majorana state (see Ref.~\cite{Haim2015})
\begin{equation}
\label{H1sf}
H=\sum_{k,\sigma} \epsilon_k \psi^{\dagger}_{k\sigma} \psi_{k\sigma} + i\gamma\sum_{k,\sigma} \left( t_\sigma\psi_{k\sigma}+t_\sigma \psi^\dagger_{k\sigma} \right) ,
\end{equation}
where the first term describes the lead, $\psi^\dagger_{k\sigma}$ being the creation operator of a spin-$\sigma$ particle with momentum $k$ and energy $\epsilon_k$, and the second term the coupling to the localised Majorana fermion $\gamma$. Without loss of generality, we assume the coupling parameters $t_\uparrow$ and $t_\downarrow$ to be real.
Using Eq.~(\ref{H1sf}) the scattering matrix in Nambu space takes the following form
\begin{equation}
\label{scattMBS}
s^{\text{M}}=
\begin{pmatrix}
r^{\text{M}}_{ee} & r^{\text{M}}_{eh} \\
r^{\text{M}}_{he} & r^{\text{M}}_{hh}
\end{pmatrix} .
\end{equation}
Here $r^{\text{M}}_{\alpha\beta}$ are matrices, in spin space, of reflection amplitudes given by
\begin{equation}
\label{Ran}
r^{\text{M}}_{ee}=\begin{pmatrix}
1 & 0 \\ 0 & 1
\end{pmatrix} +
\frac{1}{iE-\Gamma}
\begin{pmatrix}
\Gamma_\uparrow & \sqrt{\Gamma_\uparrow \Gamma_\downarrow} \\
 \sqrt{\Gamma_\uparrow \Gamma_\downarrow} & \Gamma_\downarrow 
\end{pmatrix} 
\end{equation}
and
\begin{equation}
\label{RaAn}
r^{\text{M}}_{he}=\frac{1}{iE-\Gamma}
\begin{pmatrix}
\Gamma_\uparrow & \sqrt{\Gamma_\uparrow \Gamma_\downarrow} \\
 \sqrt{\Gamma_\uparrow \Gamma_\downarrow} & \Gamma_\downarrow 
\end{pmatrix} ,
\end{equation}
where $\Gamma_\uparrow=2\pi\nu_0|t_\uparrow|^2$, $\Gamma_\downarrow=2\pi\nu_0|t_\downarrow|^2$, $\Gamma=\Gamma_\uparrow+\Gamma_\downarrow$ with $\nu_0$ being the density of states of the normal leads, while $r^{\text{M}}_{eh}$ and $r^{\text{M}}_{hh}$ are determined by  particle-hole symmetry  $r^{\text{M}}_{\alpha,\beta}(E)=[r^{\text{M}}_{-\alpha,-\beta}(-E)]^\star$.

Assuming no spin mixing to occur in the beam-splitter, the block of the  scattering matrix $s^{\text{bs}}$ for spin-$\sigma$ electrons is a $3\times 3$ unitary matrix which can be parameterized as follows\cite{Buttiker1984}
\begin{equation}
\label{Sbs}
s^{\text{bs}}_{e\sigma}=
\begin{pmatrix}
c_\sigma & \lambda_{1\sigma} & \lambda_{2\sigma} \\
\lambda_{1\sigma} & a_{1\sigma} & b_{\sigma}\\
\lambda_{2\sigma} & b_{\sigma} & a_{2\sigma}
\end {pmatrix} .
\end{equation}
Here $\lambda_{1(2)\sigma}$ is the scattering amplitude for an electron with spin $\sigma$ injected from lead $1(2)$ to be transmitted in 3 (with $\lambda_{1\sigma}^2+\lambda_{2\sigma}^2\leq 1$), $c_{\sigma}=-s_1\sqrt{1-\lambda_{1\sigma}^2-\lambda_{2\sigma}^2}$ (with $s_1=\pm 1$) is the scattering amplitude for an electron to be reflected back in lead 3, $b_{\sigma}=-\lambda_{1\sigma} \lambda_{2\sigma} (s_2+c_{\sigma}) (\lambda_{1\sigma}^2+\lambda_{2\sigma}^2)^{-1}$ (with $s_2=\pm 1$) is the amplitude for an electron  to be transmitted from lead 1 to 2, and $a_{1\sigma}=(s_2\lambda_{2\sigma}^2-c_\sigma \lambda_{1\sigma}^2) (\lambda_{1\sigma}^2+\lambda_{2\sigma}^2)^{-1}$ [$a_{2\sigma}=(s_2\lambda_{1\sigma}^2-c_\sigma \lambda_{2\sigma}^2) (\lambda_{1\sigma}^2+\lambda_{2\sigma}^2)^{-1}$] is the amplitude for an electron to be reflected back in lead $1(2)$.
Hereafter, if not otherwise stated, we fix $s_1=-1$ and $s_2=+1$.
The $3\times3$ block of the scattering matrix $s^{\text{bs}}$ for spin-$\sigma$ holes is related to the one for electrons by the electron-hole symmetry relation $s^{\text{bs}}_{h\sigma}=s^{\text{bs}\star}_{e\sigma}$, while the Andreev blocks are zero. The total scattering matrix $s^{\text{bs}}$ has dimensions $12\times12$.

\begin{figure}[t]
\centering
\includegraphics[width=\columnwidth]{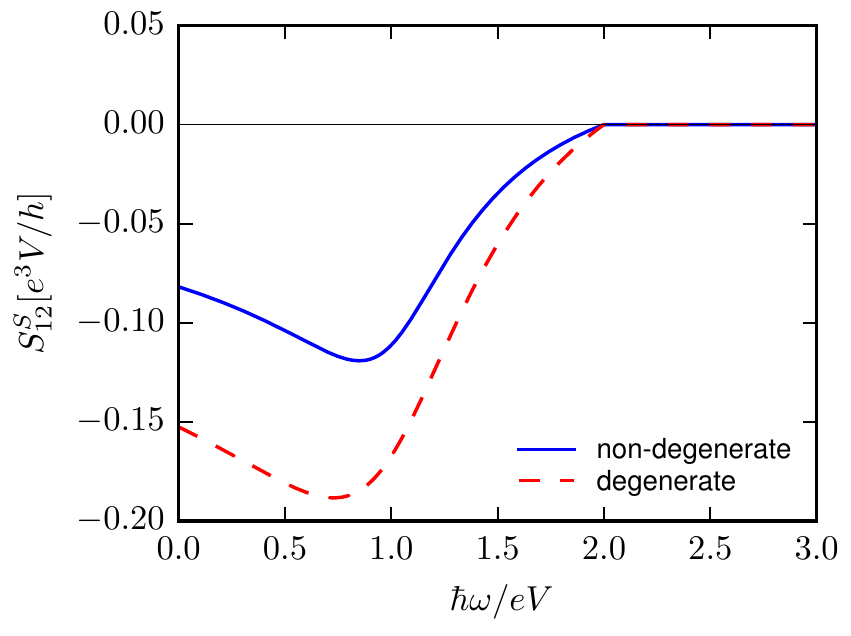}
\caption{
Symmetrized current cross correlator $S^S_{12}(\omega)$ as a function of frequency  in the presence of a MBS for a spin-degenerate system (dashed red curve) and for the case of broken spin-degeneracy (solid blue curve). For the spin degenerate case the values of the  system parameters are:  $\Gamma_\sigma=0.05 \; eV$, $\lambda_{1\sigma}=0.5\; $, $\lambda_{2\sigma}=0.6\; $, $s_1=-1$ and $s_2=+1$. For the case of broken spin degeneracy  the values of the system parameters are: $\Gamma_\uparrow=0.1 \; eV$, $\Gamma_\downarrow=0.2 \; eV$, $\lambda_{1\uparrow}=0.8$, $\lambda_{2\uparrow}=0.6$, $\lambda_{1\downarrow}=0.6$, $\lambda_{2\downarrow}=0.8$.} 
\label{Ssl}
\end{figure}

By substituting the scattering matrix obtained from the composition of $s^{\text{bs}}$ and $s^{\text{M}}$ into Eqs.~(\ref{Sspin}) and (\ref{A}) one gets the cross-correlator $S^S_{12}(\omega)$ plotted as a  dashed red line in Fig.~\ref{Ssl} for zero temperature and full spin degeneracy. 
Fig.~\ref{Ssl} shows that $S^S_{12}$ is negative at $\omega=0$ and thereafter decreases exhibiting a minimum around $\hbar\omega=eV$. 
Remarkably, the noise vanishes at frequencies higher than $2eV$.
While the occurrence of a minimum is not a generic feature, we find that $S^S_{12}(0) \le0$ and that  $S^S_{12}(\omega)=0$ for $\hbar\omega >2eV$ independently of the choice of parameters $\Gamma_\sigma$, $\lambda_{1\sigma}$ and $\lambda_{2\sigma}$, hence
these two characteristic features persist even when the spin degeneracy is removed.  As an example, the solid blue line in Fig.~\ref{Ssl} shows the  cross-correlator $S_{12}^S$  for $\Gamma_\uparrow=0.1 \; eV$, $\Gamma_\downarrow=0.2 \; eV$, $\lambda_{1\uparrow}=0.8$, $\lambda_{2\uparrow}=0.6$, $\lambda_{1\downarrow}=0.6$, $\lambda_{2\downarrow}=0.8$.

\begin{figure}[t]
\centering
\includegraphics[width=\columnwidth]{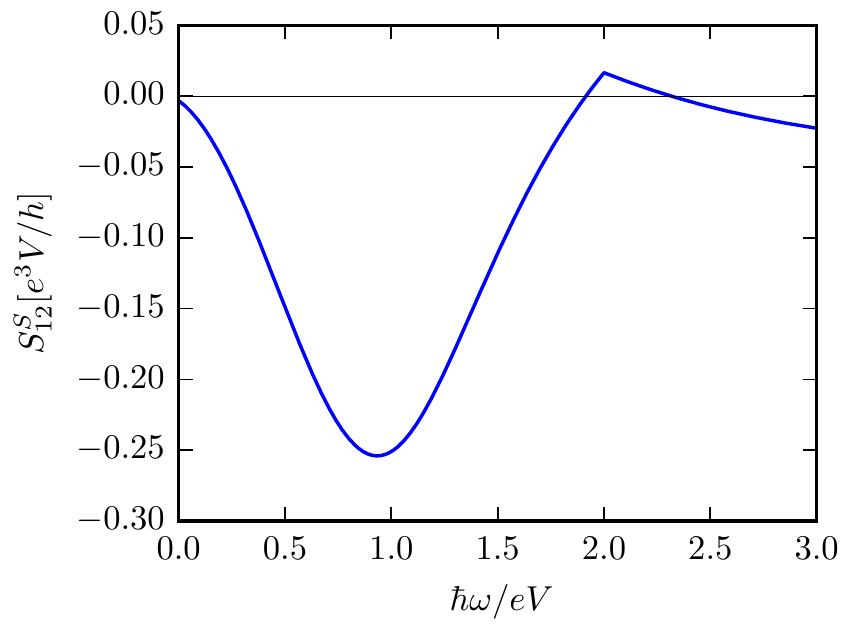}
\caption{Symmetrized current cross correlator $S^S_{12}(\omega)$ as a function of frequency  for an ABS, for the following values of the system parameters: $\Gamma_\uparrow=0.1eV$, $\Gamma_\downarrow=0.2eV$, $\lambda_{1\uparrow}=\lambda_{2\downarrow}=0.7$, $\lambda_{2\uparrow}=\lambda_{1\downarrow}=0.4$, $s_1=- 1$ and $s_2=+1$.} 
\label{SABS}
\end{figure}

In contrast to the MBS case, the behaviour of $S^S_{12}(\omega)$ is different in the presence of an ordinary zero-energy Andreev bound state (ABS). This can be realised in a Zeeman-split resonant level strongly proximized by an s-wave superconductor, which we describe through the following reflection-amplitude matrices in spin space~\cite{Haim2015}
\begin{equation}
\label{scattABS1}
r^{\text{A}}_{ee}=\frac{1}{iE-\Gamma/2}
\begin{pmatrix}
iE+\frac{\Gamma_\uparrow-\Gamma_\downarrow}{2} & 0 \\
 0 & iE-\frac{\Gamma_\uparrow-\Gamma_\downarrow}{2}
\end{pmatrix} ,
\end{equation}
\begin{equation}
\label{scattABS2}
r^{\text{A}}_{he}=\frac{\sqrt{\Gamma_\uparrow \Gamma_\downarrow}}{iE-\Gamma/2}
\begin{pmatrix}
0 & 1 \\
1 & 0
\end{pmatrix} ,
\end{equation}
where,  $r^{\text{A}}_{eh}$ and $r^{\text{A}}_{hh}$ are determined as usual by electron-hole symmetry $r^{\text{A}}_{\alpha,\beta}(E)=[r^{\text{A}}_{-\alpha,-\beta}(-E)]^\star$. We find that $S^S_{12}(\omega)$ is in general finite even at high frequencies.
As an example we plot in Fig.~\ref{SABS} the cross correlator as a function of frequency for a case of broken spin degeneracy.

As we shall show below, the vanishing cross-correlator in the presence of MBS is a result of the peculiar structural properties of the energy-dependent scattering matrix.
This is easier to understand by abandoning the beam splitter and by noticing that the index $\sigma$ in the Hamiltonian (\ref{H1sf}) can be thought of as identifying a terminal instead of a spin projection. Such a Hamiltonian would describe a Majorana state $\gamma$ separately coupled to two different single-channel, i.e. fully spin-polarized, normal terminals without explicitly needing an additional beam splitter.
In this case, the reflection amplitudes in Eqs.~(\ref{Ran}) and (\ref{RaAn}) can then be rewritten in the \textit{normal-terminal space} as
\begin{equation}
\label{RanN}
r^{\text{M}}_{ee}=\begin{pmatrix}
1 & 0 \\ 0 & 1
\end{pmatrix} +
\frac{1}{iE-\Gamma}
\begin{pmatrix}
\Gamma_1 & \sqrt{\Gamma_1 \Gamma_2} \\
 \sqrt{\Gamma_1 \Gamma_2} & \Gamma_2
\end{pmatrix} 
\end{equation}
and
\begin{equation}
\label{RaAnN}
r^{\text{M}}_{he}=\frac{1}{iE-\Gamma}
\begin{pmatrix}
\Gamma_1 & \sqrt{\Gamma_1 \Gamma_2} \\
 \sqrt{\Gamma_1 \Gamma_2} & \Gamma_2
\end{pmatrix} ,
\end{equation}
where now $\Gamma_i$ is the coupling strength to lead $i$ and $\Gamma=\Gamma_1+\Gamma_2$.
The cross-correlator can now be calculated by substituting the scattering matrix (\ref{scattMBS}), with (\ref{RanN}) and (\ref{RaAnN}), in Eq.~(\ref{Sspin}).
Assuming for the sake of simplicity $\Gamma_1=\Gamma_2=\Gamma/2$ one obtains
\begin{equation}
\label{Stermbs1}
S^{S}_{12}(\omega)=
\begin{cases}
\frac{2 e^2}{h} \frac{\Gamma^2}{|\hbar\omega|} 
\ln \frac{\Gamma^2+(|eV|-|\hbar\omega|)^2}{\Gamma^2+(eV)^2} & \mbox{if }|\hbar\omega|<2|eV|\\
0 & \mbox{if }  |\hbar\omega|\ge 2|eV|.
\end{cases}
\end{equation}
We have explicitly checked that $S^{S}_{12}(\omega)=0$ vanishes for $|\hbar\omega|>|2eV|$ also when $\Gamma_1\neq\Gamma_2$. 
Incidentally the autocorrelations for this case are given in~\ref{auto}.
As shown in~\ref{sec::vanish}, the reason why the high-frequency correlator vanishes at zero temperature, is the fact that the $2\times 2$ matrices $r_{ee}^\text{M}$ and $r_{he}^\text{M}$ are related by
\begin{equation}
\label{cond}
r_{ee}^\text{M}=\mathbb{I} +
r_{he}^\text{M},
\end{equation}
where $\mathbb{I}$ is the $2\times 2$ unity matrix.
At finite temperature $T$, the cross-correlator is exponentially suppressed for $|\hbar\omega|>2 |eV|$ as long as $k_{B}T\ll 2|eV|$ and the frequency is such that $|\hbar \omega|-2|eV|\gg k_{B}T$.
For the sake of completeness, it is important to notice that a MBS located at the end of a superconducting wire can be coupled to two normal leads through a 3-leg structure which in general does not ensure that the two leads are only coupled via the MBS.
Indeed, for the 3-leg beam splitter modelled by the scattering matrix in Eq.~(\ref{Sbs}) the two leads are not coupled separately due to the transmission coefficients $b_\sigma$.
Nevertheless, it turns out that when $s_2=+1$ the relation (\ref{cond}) holds for any choice of the other parameters and, as shown in Fig.~\ref{Ssl}, the cross-correlations vanish at high frequency.
This is however not the case when $s_2=-1$.

\section{Numerical results}
\begin{figure}[t]
\centering
\includegraphics[width=0.9\columnwidth]{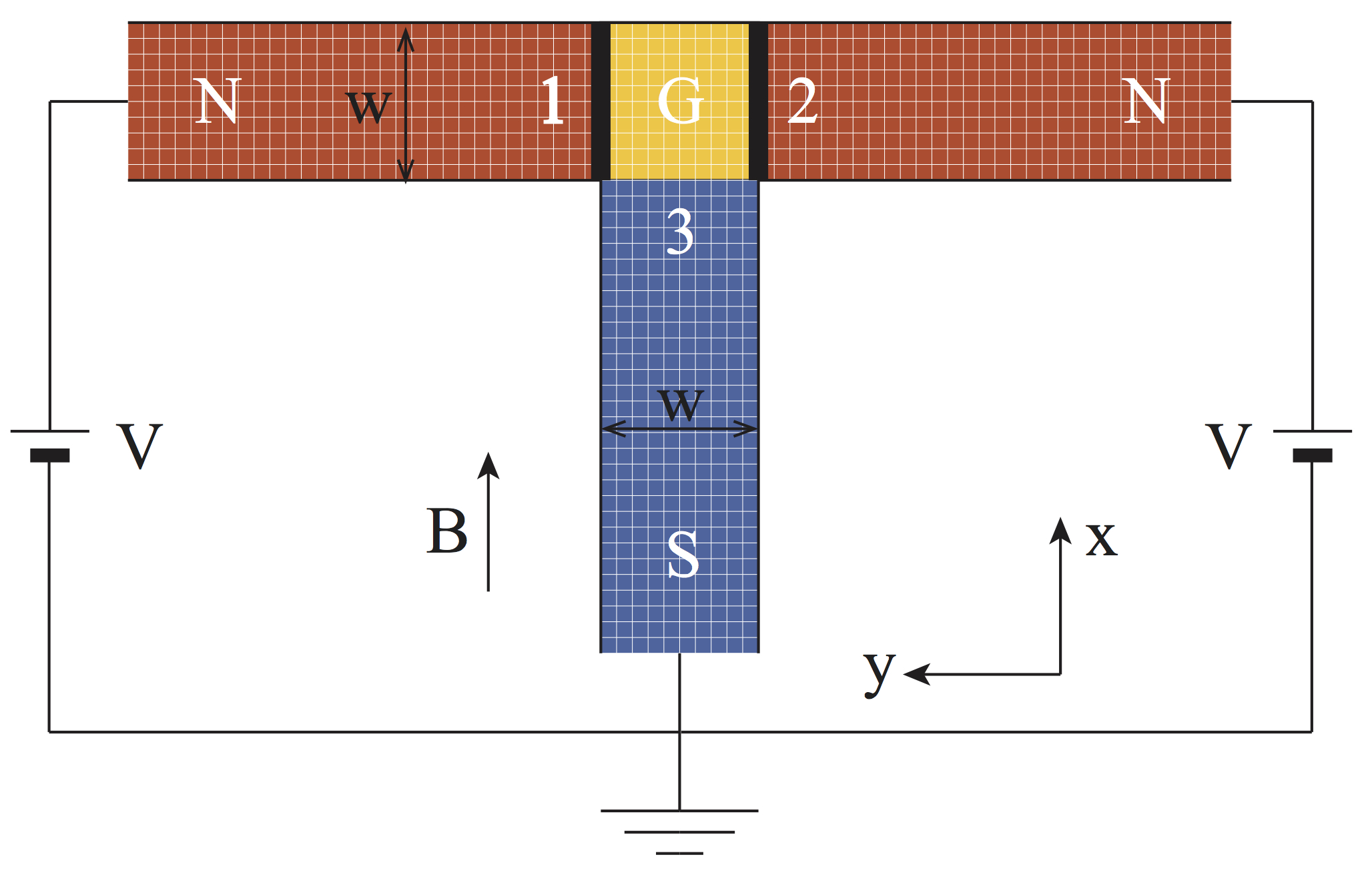}
\caption{Setup of the structure for the numerical tight-binding model. A semiconducting nanowire of width $w$ with strong spin-orbit coupling and proximized by an s-wave superconductor (vertical blue region 3) is attached to a normal semiconducting nanowire (red horizontal regions 1 and 2). Two barriers (black stripes) in proximity of the junction are used to reduce the direct coupling between the two normal leads. An in-plane magnetic field $B$, which can affect either both nanowires or just the superconducting one, controls the topological phase transition.}
\label{setupNUM}
\end{figure}

The results obtained with the minimal model reported in the previous section still hold for a more realistic situation. 
For this purpose, we consider a system based on semiconducting nanowires with a strong spin-orbit coupling.
 In the presence of a Zeeman field and an induced $s$-wave superconducting order parameter, a strongly spin-orbit coupled nanowire hosts MBS at its ends when the parameters are properly tuned (see Refs.~\cite{Lutchyn2011,Stanescu2011,Gibertini2012}).
The system we simulate is described in details in Fig.~\ref{setupNUM}:  a superconducting nanowire (vertical) is attached to the side of a normal nanowire (horizontal). Two barriers (black stripes) are included in the normal nanowire in order to suppress the direct coupling between terminals 1 and 2 and make the MBS couple separately to the two normal reservoirs, at least to some extent. 
The tight-binding Hamiltonian of the superconducting nanowire reads
\begin{align}
\label{eq:hamiltnwS}
 \cal{H}_{\rm S} &= -t\sum_{\langle i,j\rangle,\sigma} c^\dag_{i,\sigma} c_{j,\sigma} + (\varepsilon_{\rm S}-\mu)\sum_{ i,\sigma} 
c^\dag_{i,\sigma} c_{i,\sigma} \nonumber\\
& + i\lambda_{\rm{SO}} \sum_{\langle i,j\rangle,\sigma,\sigma'} (\nu'_{ij}\sigma^x_{\sigma\sigma'} - \nu_{ij}\sigma^y_{\sigma\sigma'}) 
c^\dag_{i,\sigma} c_{j,\sigma'} \\
&  + B \sum_{i,\sigma,\sigma'}\sigma^x_{\sigma\sigma'} c^\dag_{i,\sigma} c_{i,\sigma'} + \sum_{i}\left[\Delta~ c^\dag_{i,\uparrow} c^\dag_{i,\downarrow} + \text{H.c.}\right]\;,\nonumber
\end{align}
where $t$ is the hopping energy, $\varepsilon_{\rm S}= 4 t$ is a uniform on-site energy which sets the zero of energy,  $\lambda_{\text{SO}}$ is the Rashba spin-orbit (SO) 
coupling strength, $B$ is the Zeeman energy in the wire, $\Delta$ is the induced superconducting pairing, $\sigma^{i}$ are spin-1/2 Pauli matrices, $\nu_{i j} = \hat{\bm{x}} \cdot  \hat{\bm{d}}_{i j}$, and 
$\nu'_{i j} = \hat{\bm{y}} \cdot  \hat{\bm{d}}_{i j}$ with $ \hat{\bm d}_{i j} = ({\bm r}_i - {\bm r}_j)/|{\bm r}_i - {\bm r}_j|$ being the unit vector connecting 
site $j$ to site $i$.
The Hamiltonian of the normal nanowire reads
\begin{align}
\label{eq:hamiltnwN}
 {\cal H}_{\rm N} &= -t\sum_{\langle i,j\rangle,\sigma} c^\dag_{i,\sigma} c_{j,\sigma} + (\varepsilon_{\rm N}-\mu)\sum_{ i,\sigma} 
c^\dag_{i,\sigma} c_{i,\sigma}\nonumber\\
&  + B' \sum_{i,\sigma,\sigma'}\sigma^x_{\sigma\sigma'} c^\dag_{i,\sigma} c_{i,\sigma'}\;,
\end{align}
and the barriers are implemented by means of the following term: 
\begin{align}\label{eq:hamiltc}
 {\cal H}_{\rm b} &= -t \left( (\eta_{L}-1) \sum_{\langle i,j\rangle,\sigma}^{({\rm b_L})} c^\dag_{i,\sigma} c_{j,\sigma} + 
(\eta_{R}-1) \sum_{\langle i,j\rangle,\sigma}^{({\rm b_R})} c^\dag_{i,\sigma} c_{j,\sigma}  \right. \nonumber\\
&\left. +\eta_{D} \sum_{\langle i,j\rangle,\sigma}^{({\rm b_D})} c^\dag_{i,\sigma} c_{j,\sigma} \right) +
\text{H.c.} ,
\end{align}
where the superscripts $({\rm b_L})$, $({\rm b_R})$ and $({\rm b_D})$ in the  sums indicate that  the sites $i,j$ are at the interfaces of the left (L) and right (R) barriers and between the horizontal wire and the vertical one (D), respectively.
The parameters $\eta_{L}$, $\eta_{R}$ and $\eta_{D}$ are the strengths of the coupling to the left, right and bottom regions, respectively. 
Lead 3 is grounded while leads 1 and 2 are kept at the same voltage $V$.
In Eq. (\ref{eq:hamiltnwN}) $\varepsilon_{\rm N}$ is a uniform on-site energy and $B'$ is the Zeeman energy in the normal wire, which we choose to  be either 0 or the same as the Zeeman energy B of the superconducting one.
In the following, we set $\eta_D=1$, $\eta_L=\eta_R=0.1$, $\mu=0$, $\lambda_{\text{SO}}=0.1 t$, $\Delta=0.1 t$ and nanowire's width $w = 10$ sites.
Moreover, if not otherwise stated, we fix $eV=3\times 10^{-4}t$.

\begin{figure}[t]
\centering
\includegraphics[width=\columnwidth]{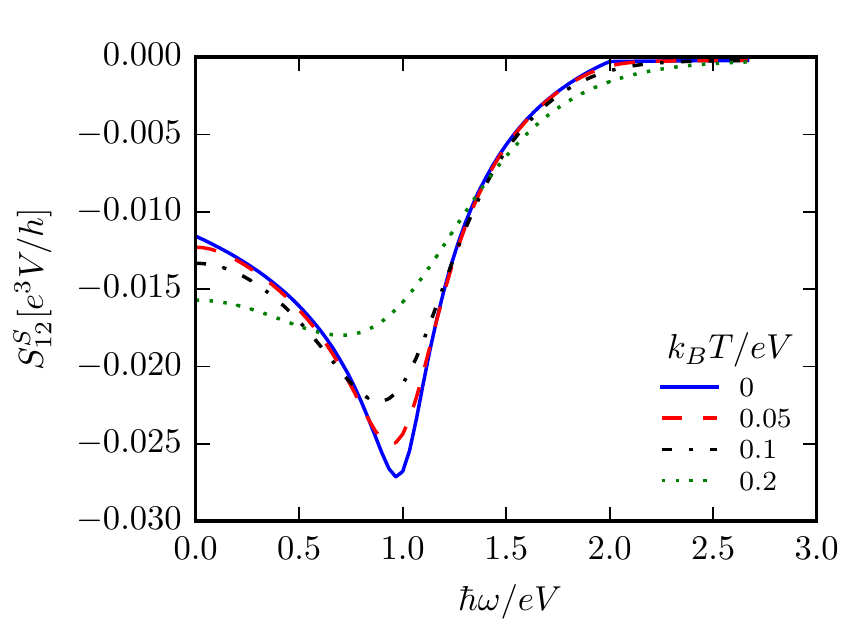}
\caption{Symmetrized noise $S^{S}_{12}(\omega)$ calculated for the realistic system in the topological phase for different temperatures, with a single open channel in the normal nanowire. The parameters for normal and superconducting nanowires are $\mu=0$, $\lambda_{\text{SO}}=0.1 t$, $B = B' = 0.2 t$, $w=10$ sites, and $\Delta=0.1 t$.} 
\label{SMbsN}
\end{figure}

Let us first consider the case when $\varepsilon_{\rm N}=\varepsilon_{\rm S}$ and $B' = B$, so that the normal nanowire supports a single, spin-polarised open channel.
We have checked that, in the topological phase ($B=0.2 t$), the Andreev and normal reflection matrices [$r^M_{he}(E)$ and $r^M_{ee}(E)$] are very well approximated by the analytical expressions (\ref{RanN}) and (\ref{RaAnN}), with $\Gamma_1\simeq \Gamma_2 \simeq 1.625 \times 10^{-5}t \simeq 0.054eV$.
In Fig.~\ref{SMbsN} the cross-correlator $S^{S}_{12}(\omega)$ is plotted for different temperatures. At zero temperature (solid blue line) the behaviour is qualitatively the same as the one for the analytical model plotted in Fig. \ref{Ssl}. At finite temperatures the dip at $\hbar \omega \simeq eV$ is suppressed, while the noise at high frequencies remains exponentially small in $\hbar\omega / k_B T$.

\begin{figure}[t]
\centering
\begin{overpic}[width=\columnwidth]{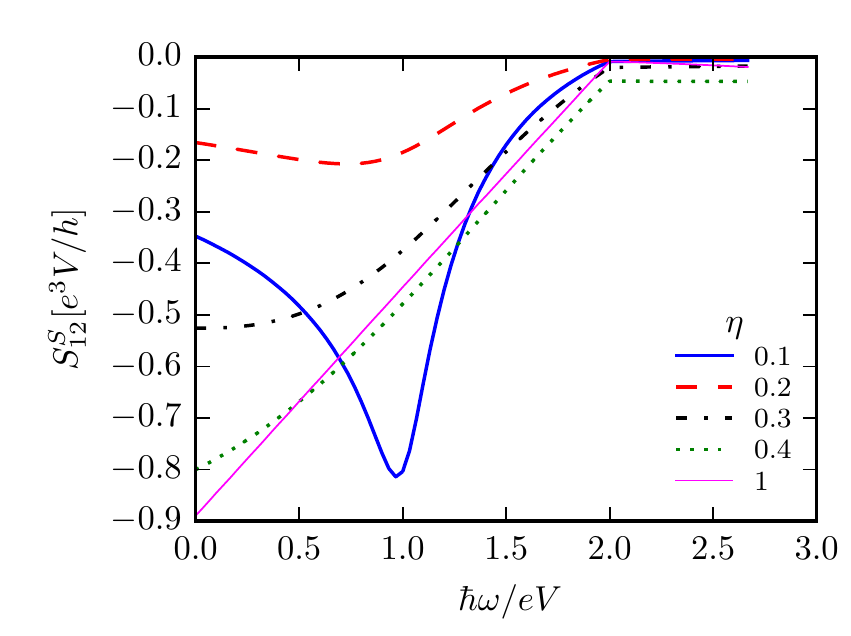}
 \put (50,25) {$\displaystyle\times 30$}
\end{overpic}
\caption{Symmetrized noise $S^{S}_{12}(\omega)$ at zero temperature calculated for the realistic system in the topological phase for different transparencies $\eta$, with a single open channel in the normal nanowire. The parameters for normal and superconducting nanowires are $\mu=0$, $\lambda_{\text{SO}}=0.1 t$, $B = B' = 0.2 t$, $w=10$ sites, and $\Delta=0.1 t$.} 
\label{SMbsTransp}
\end{figure}

As discussed at the end of Sec. \ref{a1d}, a three-leg junction gives rise in general also to a direct coupling between the two normal terminals. This coupling can be controlled by the two barriers in the normal nanowire which tune the transparencies $\eta_L$ and $\eta_R$ in Eq. (\ref{eq:hamiltc}). We studied the influence of the direct coupling on the cross-correlator $S^{S}_{12}(\omega)$ and plotted in Fig. \ref{SMbsTransp} the cross-correlator for different values of the transparencies $\eta_L$ and $\eta_R$ which, for simplicity,  we assume equal, that is $\eta_L=\eta_R=\eta$. Besides a general increase of the noise magnitude due to the broadening of the scattering probabilities peaks, one can notice that for higher transparencies $S^{S}_{12}(\omega)$ tends towards a linear behavior for small frequencies. For intermediate values of transparencies, the value of the cross-correlator at high frequencies becomes non-negligible. As expected, the direct coupling of the normal leads renders the effect of the MBS less visible.

We also investigated the role of multiple (spin-) channels in the normal nanowire by setting $\varepsilon_{\rm N}=3.8 t$ and $B'=0$, while all the other parameters are the same as in the previous case. With such a choice of parameters there are two (opposite-spin) channels for each normal terminal. 
The cross-correlator $S^{S}_{12}(\omega)$ in such a regime at zero temperature is almost equal to the $T=0$ curve in Fig. \ref{SMbsN}, {\it i.e.}, $S^{S}_{12}(0)<0$ and $S^{S}_{12}(\omega) \simeq 0$ for $|\hbar \omega |> 2| eV|$.

We shall now study the case of an ordinary ABS occurring at zero energy and verify that it yields non-vanishing cross-correlations for $|\hbar \omega |> 2| eV|$.
First we note that the occurrence of a zero-energy ABS implies a peak in the Andreev reflection probability only if at least 2 open channels are present in the normal nanowire (otherwise because of the Beri degeneracy, the Andreev reflection probability is either 1, for the topological phase, or 0, for the trivial phase).
To do so, we fix the parameters of the normal nanowire as follows: $\varepsilon_{\rm N}=3.8 t$ and $B'=0$. In order to be in the trivial phase we also put $B=0.1 t$. Furthermore, we add a gate voltage $V_g$ and a magnetic field $B''$ in the region between the barriers in order to tune the Andreev reflection probability to have just a single peak at the Fermi energy and nothing more structured in the whole subgap energy range. This is realised by an additional term in the Hamiltonian:
\begin{align}
\label{eq:hamiltG}
 {\cal H}_{\rm G} &= -V_g \sum_{ i,\sigma} 
c^\dag_{i,\sigma} c_{i,\sigma}
+ B'' \sum_{i,\sigma,\sigma'}\sigma^z_{\sigma\sigma'} c^\dag_{i,\sigma} c_{i,\sigma'}\;.
\end{align}

\begin{figure}[t]
\centering
\includegraphics[width=\columnwidth]{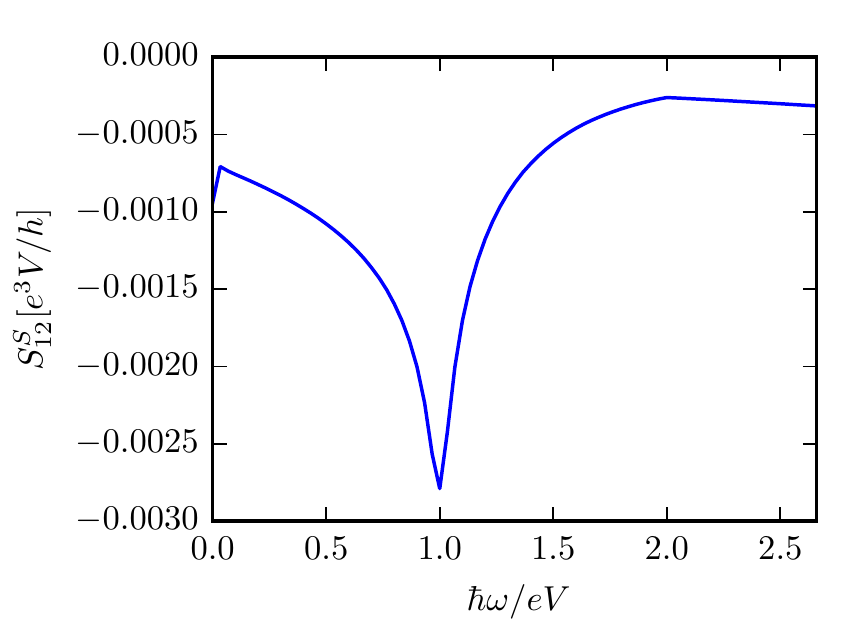}
\caption{Symmetrized noise $S^{S}_{12}(\omega)$ calculated for the realistic system in the trivial phase with a zero energy ABS at zero temperature, with two open channels in the normal nanowire. The parameters for the normal nanowire are $\mu=0$, $\varepsilon_N=3.8t$, $\lambda_{\text{SO}}=0$, $B'=0$, $w=10$ sites, while for the superconducting nanowire are $\mu=0$, $\varepsilon_S=4t$, $\lambda_{\text{SO}}=0.1 t$, $w=10$ sites, $\Delta=0.1 t$, $B=0.1 t$ and for the gate are $V_g = 0.25147785 t$ and $B'' = 0.2 t$.} 
\label{SMbs2ch}
\end{figure}

The cross-correlator $S^{S}_{12}(\omega)$ in the presence of a zero energy ABS is shown in Fig.~\ref{SMbs2ch}.
The main differences with respect to the MBS are the increasing behaviour at very small frequencies  and the evident non-zero noise at high frequencies.

\section{Conclusions}
We have presented a comprehensive study of the finite frequency current cross-correlations for a MBS coupled to two different normal reservoirs. By introducing a simple model for a MBS coupled to normal degrees of freedom combined with a three-leg beam-splitter, we have obtained all finite frequency current correlations by means of the scattering approach to mesoscopic transport. We find that the topological regime is characterised by vanishing cross correlations for $|\hbar\omega|>2|eV|$ and negative cross correlations at zero frequency. The cancellation of high-frequency fluctuations is due to the  particular relation between normal and Andreev scattering that is fulfilled in the MBS case. The absence of this feature in the noise spectrum would rule out the MBS as the origin of the vanishing cross-correlation. We have then verified that all results are still valid for a more realistic tight-binding model for  finite-width nanowires.

\section*{Acknowledgements}

This work has been supported by the EU project ``ThermiQ", by MIUR-PRIN ``Collective quantum phenomena: from strongly correlated systems to quantum simulators", by the EU project COST Action MP1209 ``Thermodynamics in the quantum regime", and by the EU project COST Action MP1201 ``NanoSC". M.G. acknowledges the hospitality of Scuola Normale Superiore, Pisa.  R.F acknowledges  the Oxford Martin School  for support and the Clarendon Laboratory for hospitality during the completion of the work.

\appendix

\section{Vanishing cross-correlations for $|\hbar\omega|>|2eV|$}
\label{sec::vanish}
In this Appendix we prove that the scattering matrices in Eqs.~(\ref{RanN}) and (\ref{RaAnN}) yield vanishing cross-correlations for frequencies $|\hbar\omega|>|2eV|$. Here we actually consider a scattering matrix that generalises the relation of Eq.~(\ref{cond}), namely including additional phases, and can thus be parametrized as 
\begin{equation}
\label{COND}
s_{\alpha \beta}^{ij}(E)=\delta_{ij}\delta_{\alpha\beta}+s_{+-}^{ij}(E)\e^{i \frac{[1-\text{sign}(\alpha)]\phi_{i}}{2}}\e^{i \frac{[1+\text{sign}(\beta)]\phi_{j}}{2}},
\end{equation}
where  $\phi_{i}$ and $\phi_{j}$ are energy-independent phases. \footnote{Particle-hole symmetry implies $\phi_i+\phi_j=-2 \text{Arg}[s^{ij}_{+-}(0)]$.}
In this Appendix, $i$ and $j$ are collective indices for transport channels which represent a pair of degrees of freedom, namely terminal number and spin.

Let us consider the case of different-channel cross-correlation. The two channels (labelled $1$ and $2$ for simplicity) can be either in the same lead with different spins or in different leads, in which case the spin is not important. Eq. \eqref{Sspin} becomes 

\begin{align}
\label{SUD_subA}
S_{12}(\omega)&=\frac{e^2}{2h} \int_{-\infty}^{+\infty} dE \sum_{\substack{\alpha,\alpha' \\ \beta,\beta'}} \sum_{l,k}
\text{sign} (\beta) \text{sign} (\beta') \\
&\times \{ \delta_{\alpha\beta} \delta_{l 1} \delta_{\alpha'\beta} \delta_{k 1} -
[s^{1l}_{\beta\alpha} (E)]^\star s^{1k}_{\beta\alpha'} (E+\hbar \omega)
\} \nonumber\\
&\times \{ \delta_{\alpha'\beta'} \delta_{k2} \delta_{\alpha\beta'} \delta_{l2} -
[s^{2k}_{\beta'\alpha'} (E+\hbar \omega)]^\star s^{2l}_{\beta'\alpha} (E)
\} 
\nonumber \\
&\times f_{\alpha}(E)\left[ 1-f_{\alpha'}(E+\hbar\omega) \right] . \nonumber
\end{align}

Let us consider the quantity
\begin{align}
\label{argSUD}
B_{\alpha\alpha'}(\omega)&= \sum_{\substack{l,k \\ \beta,\beta'}} \text{sign} (\beta) \text{sign} (\beta')
\\ \nonumber
&\times
\{ \delta_{\alpha\beta} \delta_{l1} \delta_{\alpha'\beta} \delta_{k1} -
[s^{1l}_{\beta\alpha} (E)]^\star s^{1k}_{\beta\alpha'} (E+\hbar \omega)
\}
\\ \nonumber
&\times
\{ \delta_{\alpha'\beta'} \delta_{k2} \delta_{\alpha\beta'} \delta_{l2} -
[s^{2k}_{\beta'\alpha'} (E+\hbar \omega)]^\star s^{2l}_{\beta'\alpha} (E)
\},
\end{align}
which is composed by 4 terms.

The first term
\begin{equation}
\label{B1}
B_{\alpha\alpha'}^{(1)}(\omega)=\sum_{\substack{l,k \\ \beta,\beta'}}
\text{sign} (\beta) \text{sign} (\beta')
\delta_{\alpha\beta} \delta_{l1} \delta_{\alpha'\beta} \delta_{k1} 
 \delta_{\alpha'\beta'} \delta_{k2} \delta_{\alpha\beta'} \delta_{l2} 
\end{equation}
is identically zero because of terms like $\delta_{l1} \delta_{l2}=0$.

The second term is
\begin{align}
\label{B2}
B_{\alpha\alpha'}^{(2)}(\omega)&=-\sum_{\substack{l,k\\ \beta,\beta'}}
\text{sign} (\beta) \text{sign} (\beta')
\delta_{\alpha\beta} \delta_{l1} \delta_{\alpha'\beta} \delta_{k1} 
\\ \nonumber
&\times
[s^{2k}_{\beta'\alpha'} (E+\hbar \omega)]^\star s^{2l}_{\beta'\alpha} (E)\\
\nonumber
&=-\delta_{\alpha\alpha'}\text{sign} (\alpha)\{[s^{21}_{+\alpha} (E+\hbar \omega)]^\star s^{21}_{+\alpha} (E)
\\ \nonumber
&-[s^{21}_{-\alpha} (E+\hbar \omega)]^\star s^{21}_{-\alpha} (E)\}.
\end{align}
The condition of Eq. (\ref{COND}) implies that $B_{\alpha\alpha'}^{(2)}(\omega)=0$.

The third term is
\begin{align}
\label{B3}
B_{\alpha\alpha'}^{(3)}(\omega)&=-\sum_{\substack{l,k \\ \beta,\beta'}}
\text{sign} (\beta) \text{sign} (\beta') \delta_{\alpha'\beta'} \delta_{k2} \delta_{\alpha\beta'} \delta_{l2}
\\ \nonumber
&\times
[s^{1l}_{\beta\alpha} (E)]^\star s^{1k}_{\beta\alpha'} (E+\hbar \omega)
 \\
\nonumber
&=-\delta_{\alpha\alpha'}\text{sign} (\alpha)\{[s^{12}_{+\alpha} (E)]^\star s^{12}_{+\alpha} (E+\hbar \omega)
\\ \nonumber
&-[s^{12}_{-\alpha} (E)]^\star s^{12}_{-\alpha} (E+\hbar \omega)\}.
\end{align}
As for the second term, the condition of Eq. (\ref{COND}) implies that $B_{\alpha\alpha'}^{(3)}(\omega)=0$.

The forth term is 
\begin{align}
\label{B4}
&B_{\alpha\alpha'}^{(4)}(\omega)=\sum_{\substack{l,k \\ \beta,\beta'}}
\text{sign} (\beta) \text{sign} (\beta')
[s^{1l}_{\beta\alpha} (E)]^\star 
\\ \nonumber
&\times s^{1k}_{\beta\alpha'} (E+\hbar \omega)
[s^{2k}_{\beta'\alpha'} (E+\hbar \omega)]^\star s^{2l}_{\beta'\alpha} (E) \\
\nonumber
&=\sum_{l,k}
\{[s^{1l}_{+\alpha} (E)]^\star s^{1k}_{+\alpha'} (E+\hbar \omega)-[s^{1l}_{-\alpha} (E)]^\star s^{1k}_{-\alpha'} (E+\hbar \omega)
\} \\ \nonumber
&\times \{[s^{2k}_{+\alpha'} (E+\hbar \omega)]^\star s^{2l}_{+\alpha} (E)-[s^{2k}_{-\alpha'} (E+\hbar \omega)]^\star s^{2l}_{-\alpha} (E)
\}.
\end{align}
Because of Eq. (\ref{COND}), 
the only nonzero contributions to $B_{\alpha\alpha'}^{(4)}(\omega)$ are the ones with $l=1$ and $k=2$ or $l=2$ and $k=1$,
and Eq. \eqref{B4} simplifies to
\begin{align}
B_{\alpha\alpha'}^{(4)}(\omega)&=  \text{sign}(\alpha) \text{sign}(\alpha') [(s^{12}_{+-} (E)s^{21}_{+-} (E+\hbar \omega) \e^{i (\phi_1+\phi_2)})^\star \nonumber \\
&+s^{21}_{+-} (E)s^{12}_{+-} (E+\hbar \omega) \e^{i (\phi_1+\phi_2)}]. 
\end{align}
Notice that $B_{\alpha\alpha'}^{(4)}(\omega)$ does not depend on $\alpha$ and $\alpha'$ but for the sign, thus Eq. \eqref{SUD_subA} reduces to
\begin{align}
S_{12}(\omega)&=\frac{e^2}{2h} \int_{-\infty}^{+\infty} dE [(s^{12}_{+-} (E)s^{21}_{+-} (E+\hbar \omega) \e^{i (\phi_1+\phi_2)})^\star \nonumber\\
&+s^{21}_{+-} (E)s^{12}_{+-} (E+\hbar \omega) \e^{i (\phi_1+\phi_2)}] \\
&\times \sum_{\alpha,\alpha'}   \, \text{sign}(\alpha) \text{sign}(\alpha') f_{\alpha}(E)\left[ 1-f_{\alpha'}(E+\hbar\omega) \right]. \nonumber
\end{align}

It is possible to show that for $T=0$ and $\hbar \omega > 2 eV$ or $\hbar \omega < -2 eV$ the following expression holds:
\begin{equation}
F(E,\omega) \equiv \sum_{\alpha,\alpha'}   \, \text{sign}(\alpha) \text{sign}(\alpha') f_{\alpha}(E)\left[ 1-f_{\alpha'}(E+\hbar\omega) \right] =0,
\end{equation}
thus the (symmetrized and non-symmetrized) cross-correlator is zero for frequencies higher than $2eV$.

At finite temperature $T$, the previous quantity does not vanish and evaluates to
\begin{align}
F(E,\omega)&=-\frac{1}{4} \, \text{sech}[\frac{E - eV}{2 k_B T}] \, \text{sech}[\frac{E + eV}{2 k_B T}] \,\text{sinh}[\frac{eV}{k_B T}]^2 \\ \nonumber
  &\times \text{sech}[\frac{
  E - eV + \hbar \omega}{2 k_B T}] \, \text{sech}[\frac{E + eV + \hbar \omega}{2 k_B T}] ,
\end{align}
which is exponentially vanishing, for $\hbar\omega>2eV$, as long as $k_{B}T\ll 2eV$ and the frequency is such that $\hbar\omega-2eV\gg k_{B}T$.

\section{Auto-correlations}
\label{auto}
In this Appendix we consider the auto-correlations for the model of Sec.~\ref{a1d} in the case where the beam splitter is absent (or equivalently when the scattering matrix $s^{\text{bs}}_{e\sigma}$ of Eq.~(\ref{Sbs}) is characterized by $\lambda_{1\sigma}=1$ and $\lambda_{2\sigma}=0$).
In this case the auto-correlator $S_{11}(\omega)$ for a MBS is calculated by substituting Eq.~(\ref{scattMBS}) in (\ref{Sspin}).
Assuming $t_{\uparrow}=t_{\downarrow}$ one gets:
\begin{align}
\label{S1}
& S_{11}(\omega)=\frac{e^2\Gamma^2}{2h}\int_{-\infty} ^{+\infty} dE \frac{1}{(E^2+\Gamma^2)[(E+\hbar\omega)^2+\Gamma^2]}   \times\\
& \{[4\Gamma^2+(\hbar\omega)^2]({\cal F}_{++}+{\cal F}_{--}) + (2E+\hbar\omega)^2(\cal{F}_{+-}+\cal{F}_{-+})\}  \nonumber
\end{align}
where
\begin{equation}
{\cal F}_{\alpha\beta}=f_{\alpha}(E)[1-f_{\beta}(E+\hbar\omega)] .
\end{equation}

We can discuss some limiting situations for the symmetrized noise, defined as $S_S(\omega)=S_{11}(\omega)+S_{11}(-\omega)$, and assuming $eV>0$.
At zero temperature Eq.~(\ref{S1}) gives
\begin{align}
\label{S01}
 S_S(\omega)&=\frac{2 e^2}{h} \Gamma \left[ \text{Arctan}\left(\frac{\hbar\omega+eV}{\Gamma}\right) + \text{Arctan}\left(\frac{eV}{\Gamma}\right)  \right. \nonumber \\
& +\left. \frac{\Gamma}{\hbar\omega}\ln \frac{\Gamma^2+(\hbar\omega-eV)^2}{(eV)^2+\Gamma^2} \right] ,
\end{align}
for $0<\hbar\omega<2eV$, and
\begin{equation}
\label{S02}
S_S(\omega)=\frac{2 e^2}{h} \Gamma \left[ \text{Arctan}\left(\frac{\hbar\omega+eV}{\Gamma}\right) + \text{Arctan}\left(\frac{\hbar\omega-eV}{\Gamma}\right)\right] ,
\end{equation}
for $\hbar\omega>2eV$.

At $eV=0$ one finds
\begin{equation}
S_S(\omega)=\frac{4 e^2}{h} \Gamma \text{Arctan}\left( \frac{|\hbar\omega|}{\Gamma} \right) , 
\end{equation}
whereas for $eV\neq 0$, but $\omega=0$ one finds
\begin{equation}
S_S(0)=\frac{4 e^2}{h} \Gamma \left( \text{Arctan} \frac{eV}{\Gamma} -  
\frac{\Gamma eV}{\Gamma^2+(eV)^2} \right) .
\end{equation}
The small-frequency expansion ($\hbar\omega\ll eV$ or $\hbar\omega\ll \Gamma$) of Eq.~(\ref{S01}) reads
\begin{align}
\label{S2}
S_S(\omega)\simeq S_S(0)+\frac{4 e^2}{h} \frac{\Gamma^4}{(\Gamma^2+(eV)^2)^2}\hbar\omega \\
+\frac{2 e^2}{h} \frac{\Gamma^2eV(3\Gamma^2-5(eV)^2)}{3(\Gamma^2+(eV)^2)^3}(\hbar\omega)^2  . \nonumber
\end{align}
Note that the quadratic term in $\hbar\omega$ contributes positively to the noise only if $\Gamma>\sqrt{5/3}eV$.
Now, by assuming that $\Gamma\ll eV$, i.e.  in the limit $\hbar\omega\ll\Gamma\ll eV$, Eq.~(\ref{S2}) becomes
\begin{equation}
\label{S3}
S_S(\omega)\simeq
S_S(0)+\frac{4 e^2}{h} \left( \frac{\Gamma}{eV} \right)^4 \hbar\omega ,
\end{equation}
meaning that $S_S(\omega)$ increases linearly with frequency.
On the other hand, by assuming that $\Gamma\ll\hbar\omega$, i. e. in the limit $\Gamma\ll\hbar\omega\ll eV$, Eq.~(\ref{S2}) becomes
\begin{equation}
S_S(\omega)\simeq
S_S(0)-\frac{e^2}{h}\frac{10}{3}\left( \frac{\Gamma}{eV} \right)^2 \frac{(\hbar\omega)^2}{eV} ,
\end{equation}
meaning that $S_S(\omega)$ decreases quadratically with frequency.

We notice that, for an ABS, Equations \eqref{S01} and \eqref{S02} still hold after replacing $\omega$ with $2\omega$ and $eV$ with $2eV$.

\section*{References}

\bibliography{biblio}

\end{document}